\documentclass[runningheads]{llncs}
\usepackage{graphicx}

\usepackage{amsmath}
\usepackage{amssymb}
\usepackage{multirow}
\usepackage{algorithm}
\usepackage{algorithmic}
\usepackage{color}
\begin{document}
\title{Synthesis of Contrast-Enhanced Breast MRI Using Multi-b-Value DWI-based Hierarchical Fusion Network with Attention Mechanism}
\titlerunning{Synthesis of contrast-enhanced breast MRI}
%

\author{Tianyu Zhang\inst{1,2,3, \dag} \and Luyi Han\inst{1,3,\dag} \and Anna D'Angelo\inst{4} \and Xin Wang\inst{1,2,3} \and Yuan Gao\inst{1,2,3} \and Chunyao Lu\inst{1,3} \and Jonas Teuwen\inst{5} \and Regina Beets-Tan\inst{1,2} \and Tao Tan\inst{1,6,*} \and Ritse Mann\inst{1,3}}
\authorrunning{T. Zhang et al.}
\institute{Department of Radiology, Netherlands Cancer Institute (NKI), Plesmanlaan 121, 1066 CX, Amsterdam, The Netherlands \and GROW School for Oncology and Development Biology, Maastricht University, P. O. Box 616, 6200 MD, Maastricht, The Netherlands \and Department of Diagnostic Imaging, Radboud University Medical Center, Geert Grooteplein 10, 6525 GA, Nijmegen, The Netherlands \and Dipartimento di diagnostica per immagini, Radioterapia, Oncologia ed ematologia, Fondazione Universitaria A. Gemelli, IRCCS Roma, Roma, Italy \and Department of Radiation Oncology, Netherlands Cancer Institute (NKI), Plesmanlaan 121, 1066 CX, Amsterdam, The Netherlands \and Faculty of Applied Science, Macao Polytechnic University, 999078, Macao, China \\
$\dag$ T. Z. and L. H. contributed equally to this work.\\
* Corresondence: \email{taotanjs@gmail.com}}

\maketitle              
\begin{abstract}
Magnetic resonance imaging (MRI) is the most sensitive technique for breast cancer detection among current clinical imaging modalities. Contrast-enhanced MRI (CE-MRI)  provides superior differentiation between tumors and invaded healthy tissue, and has become an indispensable technique in the detection and evaluation of cancer. However, the use of gadolinium-based contrast agents (GBCA) to obtain CE-MRI may be associated with nephrogenic systemic fibrosis and may lead to bioaccumulation in the brain, posing a potential risk to human health. Moreover, and likely more important, the use of gadolinium-based contrast agents requires the cannulation of a vein, and the injection of the contrast media which is cumbersome and places a burden on the patient. To reduce the use of contrast agents, diffusion-weighted imaging (DWI) is emerging as a key imaging technique, although currently usually complementing breast CE-MRI. In this study, we develop a multi-sequence fusion network to synthesize CE-MRI based on T1-weighted MRI and DWIs. DWIs with different b-values are fused to efficiently utilize the difference features of DWIs. Rather than proposing a pure data-driven approach, we invent a multi-sequence attention module to obtain refined feature maps, and leverage hierarchical representation information fused at different scales while utilizing the contributions from different sequences from a model-driven approach by introducing the weighted difference module. The results show that the multi-b-value DWI-based fusion model can potentially be used to synthesize CE-MRI, thus theoretically reducing or avoiding the use of GBCA, thereby minimizing the burden to patients. Our code is available at \url{https://github.com/Netherlands-Cancer-Institute/CE-MRI}.
\end{abstract}


\section{Introduction}
Breast cancer is the most common cancer and the leading cause of cancer death in women~\cite{sung2021global}. Early detection of breast cancer allows patients to receive timely treatment, which may have less burden and a higher probability of survival~\cite{goldhirsch2013personalizing}. Among current clinical imaging modalities, magnetic resonance imaging (MRI) has the highest sensitivity for breast cancer detection~\cite{mann2019breast}. Especially, contrast-enhanced MRI (CE-MRI) can identify tumors well and has become an indispensable technique for detecting and defining cancer~\cite{mann2019contrast}. However, the use of gadolinium-based contrast agents (GBCA) requires iv-cannulation, which is a burden to patients, time consuming and cumbersome in a screening situation. Moreover, contrast administration can lead to allergic reactions and finaly CE-MRI may be associated with nephrogenic systemic fibrosis and lead to bioaccumulation in the brain, posing a potential risk to human health~\cite{marckmann2006nephrogenic,broome2007gadodiamide,kanda2014high,olchowy2017presence,nguyen2020dentate}. In 2017, the European Medicines Agency concluded its review of GBCA, confirming recommendations to restrict the use of certain linear GBCA used in MRI body scans and to suspend the authorization of other contrast agents, albeit macrocyclic agents can still be freely used~\cite{kleesiek2019can}.

With the development of computer technology, artificial intelligence-based methods have shown potential in image generation and have received extensive attention. Some studies have shown that some generative models can effectively perform mutual synthesis between MR, CT, and PET~\cite{yi2019generative}. Among them, synthesis of CE-MRI is very important as mentioned above, but few studies have been done by researchers in this area due to its challenging nature. Li et al. analyzed and studied the feasibility of using T1-weighted MRI and T2-weighted MRI to synthesize CE-MRI based on deep learning model~\cite{li2022virtual}. Their results showed that the model they developed could potentially synthesize CE-MRI and outperform other cohort models. However, MRI source data of too few sequences (only T1 and T2) may not provide enough valuable informative to effectively synthesize CE-MRI. In another study, Chung et al. investigated the feasibility of using deep learning (a simple U-Net structure) to simulate contrast-enhanced breast MRI of invasive breast cancer, using source data including T1-weighted non-fat-suppressed MRI, T1-weighted fat-suppressed MRI, T2-weighted fat-suppressed MRI, DWI, and apparent diffusion coefficient~\cite{chung2022deep}. However, obtaining a complete MRI sequence makes the examination costly and time-consuming. On the other hand, the information provided by multi-sequences may be redundant and may not contain the relevant information of CE-MRI. Therefore, it is necessary to focus on the most promising sequences to synthesize CE-MRI.

Diffusion-weighted imaging (DWI) is emerging as a key imaging technique to complement breast CE-MRI~\cite{baltzer2020diffusion}. DWI can provide information on cell density and tissue microstructure based on the diffusion of tissue water. Studies have shown that DWI could be used to detect lesions, distinguish malignant from benign breast lesions, predict patient prognosis, etc~\cite{baltzer2020diffusion,van2021factors,iima2020diffusion,amornsiripanitch2019diffusion,partridge2019diffusion}. In particular, DWI can capture the dynamic diffusion state of water molecules to estimate the vascular distribution in tissues, which is closely related to the contrast-enhanced regions in CE-MRI. DWI may be a valuable alternative in breast cancer detection in patients with contraindications to GBCA~\cite{baltzer2020diffusion}. Inspired by this, we develop a multi-sequence fusion network based on T1-weighted MRI and multi-b-value DWI to synthesize CE-MRI. Our contributions are as follows:
\begin{itemize}
    \item[i] From the perspective of method, we innovatively proposed a multi-sequence fusion model, designed for combining T1-weighted imaging and multi-b-value DWI to synthesize CE-MRI for the first time.
    \item[ii] We invented hierarchical fusion module, weighted difference module and multi-sequence attention module to enhance the fusion at different scale, to control the contribution of different sequence and maximising the usage of the information within and across sequences.
    \item[iii] From the perspective of clinical application, our proposed model can be used to synthesize CE-MRI, which is expected to reduce the use of GBCA.
%
\end{itemize}

\section{Methods}
\subsection{Patient collection and pre-processing}
This study was approved by Institutional Review Board of our cancer institute with a waiver of informed consent. We retrospectively collected 765 patients with breast cancer presenting at our cancer institute from January 2015 to November 2020, all patients had biopsy-proven breast cancers (all cancers included in this study were invasive breast cancers, and ductal carcinoma in situ had been excluded). The MRIs were acquired with Philips Ingenia 3.0-T scanners, and overall, three sequences were present in the in-house dataset, including T1-weighted fat-suppressed MRI, contrast-enhanced T1-weighted MRI and DWI. DWI consists of 4 different b-values (b=0, b=150, b=800 and b=1500). All MRIs were resampled to 1 mm isotropic voxels and uniformly sized, resulting in volumes of 352×352 pixel images with 176 slices per MRI, and subsequent registration was performed based on Advanced Normalization Tools (ANTs)~\cite{avants2011reproducible}.

\subsection{Model}
\begin{figure}[t]
\includegraphics[width=\textwidth]{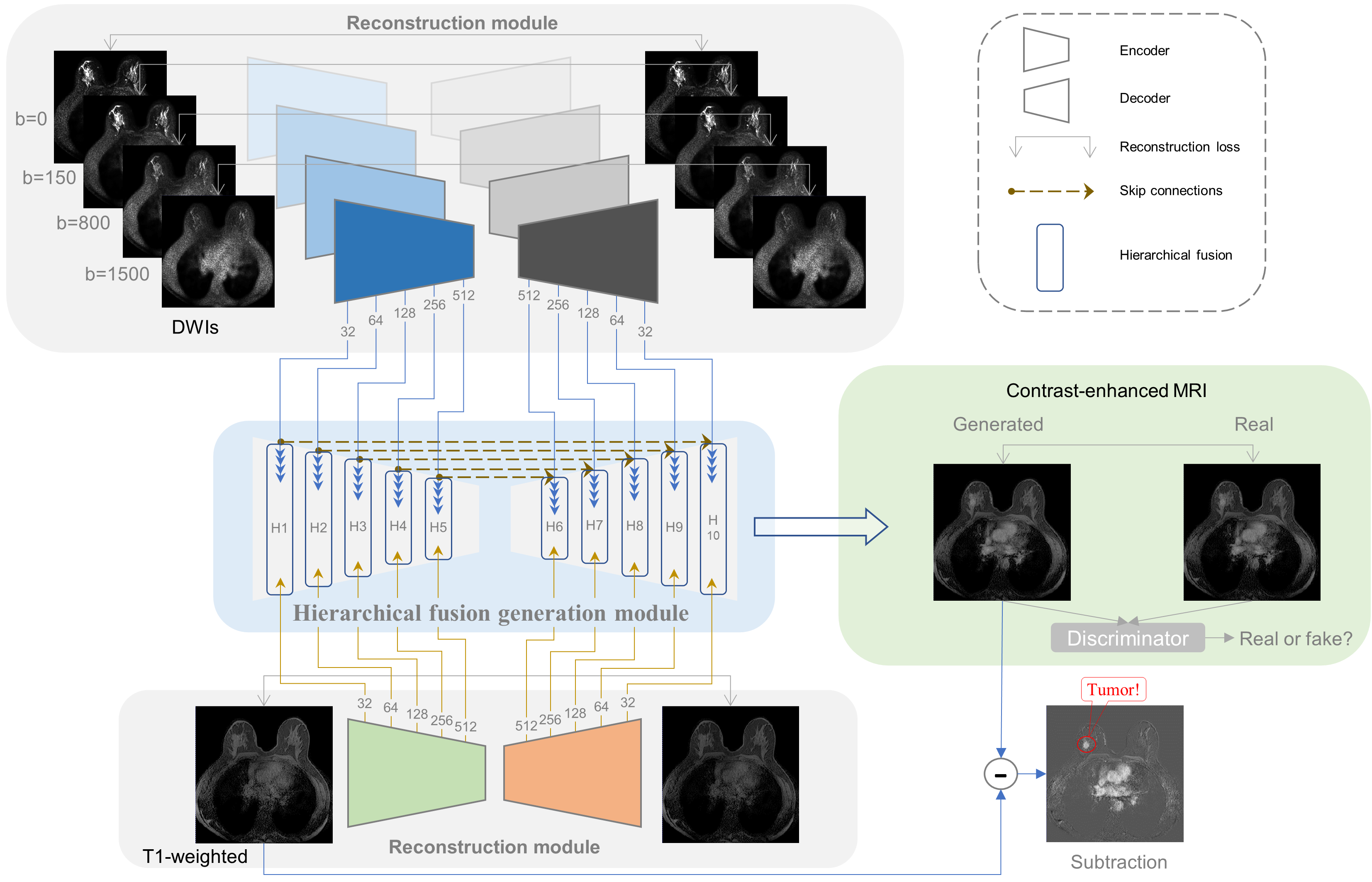}
\caption{Model details and flowchart for this study.} \label{fig1}
\end{figure}

Figure.~\ref{fig1} illustrates the structure of the proposed model. First, the reconstruction module is used to automatically encode and decode each input MRI sequence information to obtain the latent representation of different MRI sequences at multi-scale levels. Then, the hierarchical fusion module is used to extract the hierarchical representation information and fuse them at different scales. 

In each convolutional layer group of the reconstruction module, we use two 3 $\times$ 3 filters (same padding) with strides 1 and 2, respectively. The filters are followed by batch normalization, and after batch normalization, the activation functions \textit{LeakyReLU} (with a slope of 0.2) and \textit{ReLU} are used in the encoder and decoder, respectively. The $l_1$-norm is used as a reconstruction loss to measure the difference between the reconstructed image and the ground truth.

\begin{figure}[t]
\includegraphics[width=\textwidth]{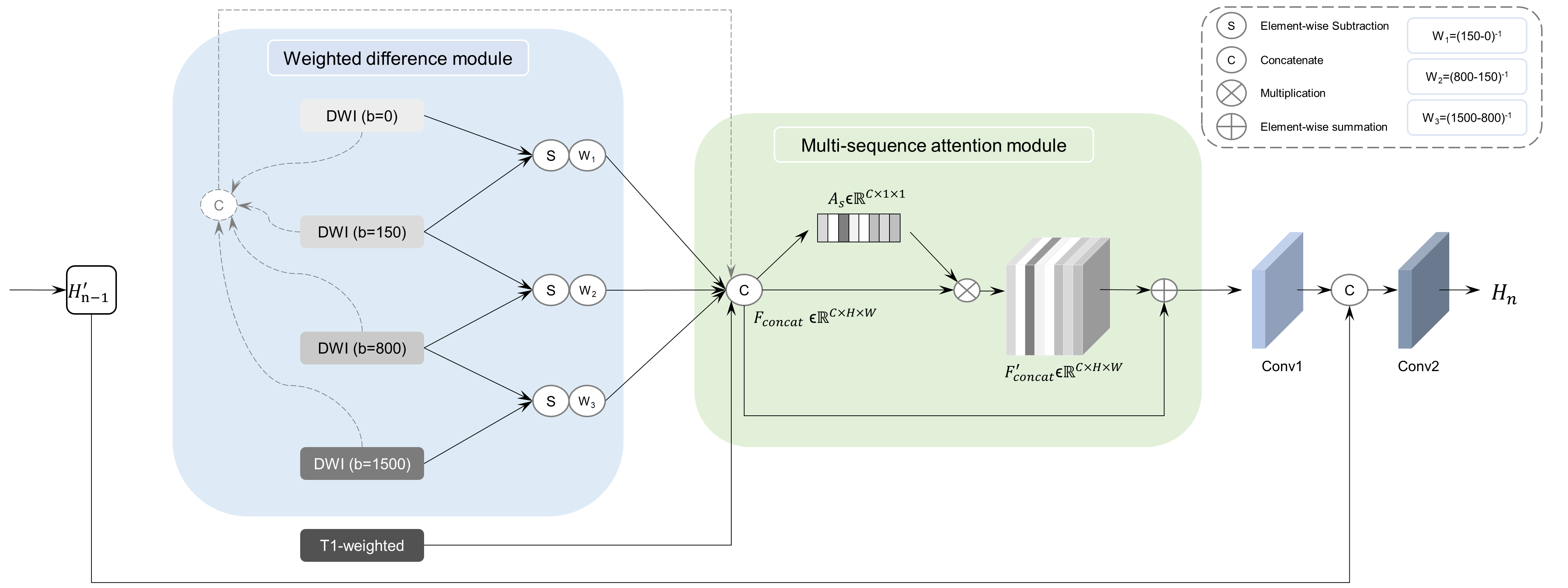}
\caption{Detailed structure of the hierarchical fusion module.} \label{fig2}
\end{figure}

Figure.~\ref{fig2} shows the detailed structure of the hierarchical fusion module, which includes two sub-modules, a weighted difference module and a multi-sequence attention module. The calculation of the apparent diffusion coefficient (ADC) map is shown in Eq.~\ref{eq:adc}, which provides a quantitative measure of observed diffusion restriction in DWIs. Inspired by ADC, a weighted difference module is designed, in which the neural network is used to simulate the dynamic analysis of the $\ln$ function, and the element-wise subtraction algorithm is used to extract the differentiation features between DWIs with different b-values, and finally the features are weighted to obtain weighted feature maps ($F_{\text{DWI}}$, Eq.~\ref{eq:wdm}).

\begin{equation}
    \label{eq:adc}
    \begin{aligned}
        \text{ADC}=-\ln{({S_{h}}/{S_{l}})}/({b_{h}}-{b_{l}})=[\ln{({S_{l}})}-\ln{({S_{h}})}]/({b_{h}}-{b_{l}})
    \end{aligned}
\end{equation}

\begin{equation}
    \label{eq:wdm}
    \begin{aligned}
        {F_{\text{DWI}}}=[{f_{\theta^{l}}}(S_{l})-{f_{\theta^{h}}}(S_{h})]/({b_{h}}-{b_{l}})  \\
    \end{aligned}
\end{equation}
where $S_{l}$ and $S_{h}$ represent the image signals obtained from lower b value $b_{l}$ and higher $b_{h}$, $f_{\theta^{l}}$ and $f_{\theta^{h}}$ represent the corresponding neural networks for DWI with a lower and higher b value.

In the multi-sequence attention module, a channel-based attention mechanism is designed to automatically apply weights (${A}_{s}$) to feature maps ($F_{concat}$) from different sequences to obtain a refined feature map ($F_{concat}^{'}$), as shown in Eq.~\ref{eq:msa}. The input feature maps ($F_{concat}$) go through the maximum pooling layer and the average pooling layer respectively, and then are added element-wise after passing through the shared fully connected neural network, and finally the weight map ${A}_{s}$ is generated after passing through the activation function, as shown in  Eq.~\ref{eq:msa2}.

\begin{equation}
    \label{eq:msa}
    \begin{aligned}
        {F_{concat}^{'}\in{\mathbb{R}^{{C}\times{H}\times{W}}}}={F_{concat}}\otimes{{A}_{s}}  \\
    \end{aligned}
\end{equation}
\begin{equation}
    \label{eq:msa2}
    \begin{aligned}
        {{A}_{s}}={\sigma({f_{\theta^{fc}}}({AvgPool({F_{concat}})}})\oplus{f_{\theta^{fc}}}({MaxPool}({F_{concat}})))
    \end{aligned}
\end{equation}

where $\otimes$ represents element-wise multiplication, $\oplus$ represents element-wise summation, $\sigma$ represents the sigmoid function, $\theta^{fc}$ represents the corresponding network parameters of the shared fully-connected neural network, and $AvgPool$ and $MaxPool$ represent average pooling and maximum pooling operations, respectively. 

In the synthesis process, the generator $\mathcal{G}$ tries to generate an image according to the input multi-sequence MRI ($d_1$, $d_2$, $d_3$, $d_4$, $t_1$), and the discriminator $\mathcal{D}$ tries to distinguish the generated image ${G}$($d_1$, $d_2$, $d_3$, $d_4$, $t_1$) from the real image $y$, and at the same time, the generator tries to generate a realistic image to mislead the discriminator. The generator's objective function is as follows:
\begin{equation}
\label{eq:LG}
\begin{aligned}
        \mathcal{L}_{G}(G,D)=&\Bbb{E}_{{d_1},{d_2},{d_3},{d_4},{t_1}{\sim}{{pro}_{data}({d_1},{d_2},{d_3},{d_4},{t_1})}}\\
        &[\log(1-{D({d_1},{d_2},{d_3},{d_4},{t_1},({G}({d_1},{d_2},{d_3},{d_4},{t_1})))})] \\
        &+\lambda_{1}\Bbb{E}_{{d_1},{d_2},{d_3},{d_4},{t_1},y}[{\|y-{G}({d_1},{d_2},{d_3},{d_4},{t_1})\|_1}]
\end{aligned}
\end{equation}
and the discriminator's objective function is as follows:

\begin{equation}
    \label{eq:LD}
    \begin{aligned}
        \mathcal{L}_{D}(G,D)=&\Bbb{E}_{y{\sim}{{pro}_{data}(y)}}[\log{D(y)]}\\
        &+\Bbb{E}_{{d_1},{d_2},{d_3},{d_4},{t_1}{\sim}{{pro}_{data}({d_1},{d_2},{d_3},{d_4},{d_1})}}\\
        &[\log{(1-{D({G}({d_1},{d_2},{d_3},{d_4},{t_1}))})}]
    \end{aligned}
\end{equation}
where $pro_{data}(d_1,d_2,d_3, d_4, t_1)$ represents the empirical joint distribution of inputs $d_1$ ($DWI_{b0}$), $d_2$ ($DWI_{b150}$), $d_3$ ($DWI_{b800}$), $d_4$ ($DWI_{b1500}$) and $t_1$ (T1-weighted MRI), $\lambda_1$ is a non-negative trade-off parameter, and $l_1$-norm is used to measure the difference between the generated image and the corresponding ground truth. The architecture of the discriminator includes five convolutional layers, and in each convolutional layer, 3 $\times$ 3 filters with stride 2 are used. Each filter is followed by batch normalization, and after batch normalization, the activation function \textit{LeakyReLU} (with a slope of 0.2) is used. The numbers of filters are 32, 64, 128, 256 and 512, respectively.

\subsection{Visualization}
The T1-weighted images and the contrast-enhanced images were subtracted to obtain a difference MRI to clearly reveal the enhanced regions in the CE-MRI.

\subsection{Experiment settings}
Based on the ratio of 8:2, the training set and independent test set of the in-house dataset have 612 and 153 cases, respectively. The trade-off parameter $\lambda_1$ was set to 100 during training, and the trade-off parameter of the reconstruction loss in the reconstruction module is set to 5. Masks for all breasts were used (weighted by a factor of 100 during the calculation of the loss between generated and real CE-MRI) to reduce the influence of signals in the thoracic area. The batch was set to 8 for 100 epochs, the initial learning rate was 1e-3 with a decay factor of 0.8 every 5 epochs (total run time is about 60 hours). $Adam$ optimizer was applied to update the model parameters. MMgSN-Net~\cite{li2022virtual} and the method of Chung et al.~\cite{chung2022deep} were used as cohort models, and all models were trained on NVIDIA RTX A6000 48 GB GPU.
\subsection{Evaluation metrics}
Results analysis was performed by Python 3.7. Structural Similarity Index Measurement (SSIM), Peak Signal-to-Noise Ratio (PSNR) and Normalized Mean Squared Error (NMSE) were used as metrics, all formulas as follows:
\begin{equation}
    \label{eq:ssim}
    \begin{aligned}
        {SSIM}=\frac{(2\mu_{y(x)}\mu_{G(x)}+{c}_{1})(2\sigma_{y(x)G(x)}+{c}_{2})}{({\mu_{y(x)}^{2}}+{\mu_{G(x)}^{2}}+{c}_{1})({\sigma_{y(x)}^{2}}+{\sigma_{G(x)}^{2}}+{c}_{2})}
    \end{aligned}
\end{equation}

\begin{equation}
    \label{eq:psnr}
    \begin{aligned}
        {PSNR}=10\log_{10}\frac{\max^{2}(y(x), G(x))}{\frac{1}{N}\|y(x)-G(x)\|_{2}^{2}}
    \end{aligned}
\end{equation}

\begin{equation}
    \label{eq:nmse}
    \begin{aligned}
        {NMSE}=\frac{y(x)-G(x)}{\|y(x)\|_{2}^{2}}
    \end{aligned}
\end{equation}

where ${G}(x)$ represents a generated image, $y(x)$ represents a ground-truth image, $\mu_{y(x)}$ and $\mu_{{G}(x)}$ represent the mean of $y(x)$ and $G(x)$, respectively, $\sigma_{y(x)}$ and $\sigma_{G(x)}$ represent the variance of $y(x)$ and ${G}(x)$, respectively, $\sigma_{{y(x)}{G}(x)}$ represents the covariance of $y(x)$ and ${G}(x)$, and $c_1$ and $c_2$ represent positive constants used to avoid null denominators.

\section{Results}
First, we compare the performance of different existing methods on synthetic CE-MRI using our source data, The quantitative indicators used include PSNR, SSIM and NMSE. As shown in Table ~\ref{table1}, the SSIM of MMgSN-Net~\cite{li2022virtual} and the method of Chung et al.~\cite{chung2022deep} in synthesizing ceT1 MRI is 86.61 $\pm$ 2.52 and 87.58 $\pm$ 2.68, respectively, the PSNR is 26.39 $\pm$ 1.38 and 27.80 $\pm$ 1.56, respectively, and the NMSE is 0.0982 $\pm$ 0.038 and 0.0692 $\pm$ 0.035, respectively. In contrast, our proposed multi-sequence fusion model achieves better SSIM of 89.93 $\pm$ 2.91, better PSNR of 28.92 $\pm$ 1.63 and better NMSE of 0.0585 $\pm$ 0.026 in synthesizing ceT1 MRI, outperforming existing cohort models.

MMgSN-Net~\cite{li2022virtual} combined T1-weighted and T2-weighted MRI in their work to synthesize CE-MRI. Here we combined T1-weighted MRI and DWI with a b-value of 0 according to their method, but the model did not perform well. It may be because their model can only combine bi-modality and cannot integrate the features of all sequences, so it cannot mine the difference features between multiple b-values, which limits the performance of the model. In addition, although the method of Chung et al.~\cite{chung2022deep} used full-sequence MRI to synthesize CE-MRI, it would be advantageous to obtain synthetic CE-MRI images using as little data as possible, taking advantage of the most contributing sequences. They did not take advantage of multi-b-value DWI, nor did they use the hierarchical fusion module to fully fuse the hierarchical features of multi-sequence MRI.

\begin{table}[t]
\centering
\caption{Results for synthesizing breast ceT1 MRI for different models.}
\label{table1}
\setlength{\tabcolsep}{3pt}
\begin{tabular}{llcccc}
\hline 
Method & Sequence & SSIM$\uparrow$ & PSNR$\uparrow$ & NMSE$\downarrow$ &\\ \hline
MMgSN-Net~\cite{li2022virtual} & T1+DWI(0) &86.61 $\pm$ 2.52 & 26.39 $\pm$ 1.38 &0.0982 $\pm$ 0.038\\
Chung et al.~\cite{chung2022deep} & T1+DWIs &87.58 $\pm$ 2.68 & 27.80 $\pm$ 1.56 &0.0692 $\pm$ 0.035 \\
\textbf{Proposed} & \textbf{T1+DWIs} &\textbf{89.93 $\pm$ 2.91} & \textbf{28.92 $\pm$ 1.63} &\textbf{0.0585 $\pm$ 0.026}\\
 \hline
\end{tabular}
\end{table}

As described in Methods, the proposed model consists of several key components, including a hierarchical fusion generation module, a weighted difference module, and a multi-sequence attention module. Therefore, ablation studies were performed to demonstrate the importance and effectiveness of our three key components. Several network structures were selected for comparison, as follows: (1) Input-level fusion network without other modules (called IF-Net), (2) Hierarchical fusion generation network combined with reconstruction module, without weighted difference module and multi-sequence attention module (called HF-Net), (3) Hierarchical fusion generation network with weighted difference module (called HFWD-net), (4) Hierarchical fusion generation network with weighted difference module and multi-sequence attention module (proposed model). As shown in Table ~\ref{table2}, IF-Net achieves SSIM of 87.25 $\pm$ 2.62 and PSNR of 26.51 $\pm$ 1.52 in synthesizing ceT1 MRI. HF-Net achieves SSIM of 88.32 $\pm$ 2.70 and PSNR of 27.95 $\pm$ 1.59 in synthesizing ceT1 MRI. After adding the weighted difference module, SSIM and PSNR were improved to 89.18 $\pm$ 2.73 and 28.45 $\pm$ 1.61, respectively. Finally, the addition of the multi-sequence attention module further improved the performance of the model, with SSIM of 89.93 $\pm$ 2.91, PSNR of 28.92 $\pm$ 1.63, and NMSE of 0.0585 $\pm$ 0.026.

\begin{table}[htbp]
\centering
\caption{Ablation results for synthesizing breast ceT1 MRI.}
\label{table2}
\setlength{\tabcolsep}{3pt}
\begin{tabular}{llcccc}
\hline 
Method & Sequence & SSIM$\uparrow$ & PSNR$\uparrow$ & NMSE$\downarrow$ &\\ \hline
IF-Net & T1+DWIs &87.25 $\pm$ 2.62 & 26.51 $\pm$ 1.52 &0.0722 $\pm$ 0.032 \\
HF-Net & T1+DWIs &88.32 $\pm$ 2.70 & 27.95 $\pm$ 1.59 &0.0671 $\pm$ 0.033 \\
HFWD-Net & T1+DWIs &89.18 $\pm$ 2.73 & 28.45 $\pm$ 1.61 &0.0622 $\pm$ 0.030 \\
\textbf{Proposed} & \textbf{T1+DWIs} &\textbf{89.93 $\pm$ 2.91} & \textbf{28.92 $\pm$ 1.63} &\textbf{0.0585 $\pm$ 0.026}\\
 \hline
\end{tabular}
\end{table}

\begin{figure}[t]
\includegraphics[width=\textwidth]{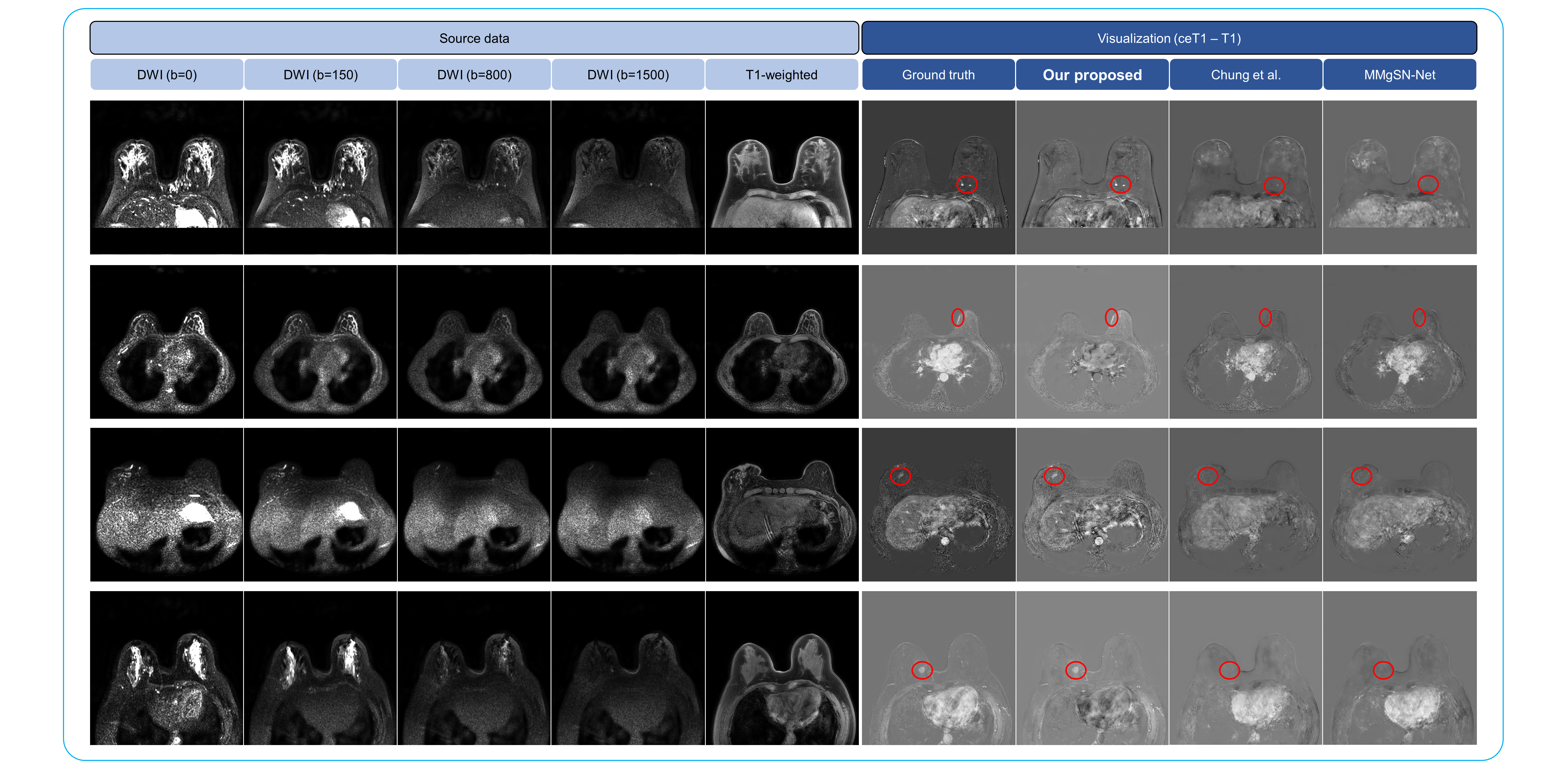}
\caption{Some examples of source data and visualization results} 
\label{fig3}
\end{figure}

The visualization results of random samples are shown in Figure.~\ref{fig3}. It can be seen from the visualization results that after the difference between the generated CE-MRI and the original T1-weighted MRI, the lesion position of the breast is highlighted, which is the same as the real enhanced position. See Supplementary Material for more visualization results, including visualizations of breast CE-MRI synthesized in axial, coronal, and sagittal planes.

\section{Conclusion}
We have developed a multi-sequence fusion network based on multi-b-value DWI to synthesize CE-MRI, using source data including DWIs and T1-weighted fat-suppressed MRI. Compared to existing methods, we avoid the challenges of using full-sequence MRI and aim to be selective on valuable source data DWI. Hierarchical fusion generation module, weighted difference module, and multi-sequence attention module have all been shown to improve the performance of synthesizing target images by addressing the problems of synthesis at different scales, leveraging differentiable information within and across sequences. Given that current research on synthetic CE-MRI is relatively sparse and challenging, our study provides a novel approach that may be instructive for future research based on DWIs. Our further work will be to conduct reader studies to verify the clinical value of our research in downstream applications, such as helping radiologists on detecting tumors. Our proposed model can potentially be used to synthesize CE-MRI, which is expected to reduce or avoid the use of GBCA, thereby optimizing logistics and minimizing potential risks to patients.
%

%

\bibliographystyle{splncs04}
\bibliography{egbib}
\end{document}